\documentclass[runningheads]{llncs}
\usepackage{graphicx}

\usepackage{booktabs}
\usepackage{color}
\usepackage{array}
\usepackage{hyperref}
\usepackage{ragged2e}
\usepackage{enumitem}

\usepackage[utf8]{inputenc}
\usepackage{csquotes}
\usepackage[super]{nth}
 \usepackage[switch]{lineno} 

\usepackage[colorinlistoftodos,prependcaption]{todonotes}
\hypersetup{
    colorlinks,
    citecolor=blue,
    filecolor=blue,
    linkcolor=black,
    urlcolor=black
}

\newcolumntype{L}[1]{>{\raggedright\let\newline\\\arraybackslash\hspace{0pt}}m{#1}}
\newcolumntype{C}[1]{>{\centering\let\newline\\\arraybackslash\hspace{0pt}}m{#1}}

\newcounter{EKXCommentsCounter}

\begin{document}
\title{Continuous Software Engineering in the Wild}
%
%
\author{Eriks Klotins\inst{1}\orcidID{0000-0002-1987-2234} \and
Tony Gorschek\inst{1,2}\orcidID{0000-0002-3646-235X}}
\authorrunning{E. Klotins et al.}
%
\institute{Software Engineering Research Lab (SERL)\\Blekinge Institute of Technology\\
\email{eriks.klotins@bth.se}, \email{tony.gorschek@bth.se}
\and
fortiss GmbH, Germany
}


%
\maketitle              
\begin{abstract}
Software is becoming a critical component of most products and organizational functions. The ability to continuously improve software determines how well the organization can respond to market opportunities. Continuous software engineering promises numerous advantages over sprint-based or plan-driven development. However, implementing a continuous software engineering pipeline in an existing organization is challenging.

In this invited position paper, we discuss the adoption challenges and argue for a more systematic methodology to drive the adoption of continuous engineering. Our discussion is based on ongoing work with several industrial partners as well as experience reported in both state-of-practice and state-of-the-art.

We conclude that the adoption of continuous software engineering primarily requires analysis of the organization, its goals, and constraints. One size does not fit all purposes, meaning that many of the principles behind continuous engineering are relevant for most organizations, but the level of realization and the benefits may still vary. The main hindrances to continuous flow of software arise from sub-optimal organizational structures and the lack of alignment. Once those are removed, the organization can implement automation to further improve the software delivery.

\end{abstract}

\section{Introduction}

Software is a critical component of most products, services, manufacturing processes, and back-office functions. The ability to continuously improve software is crucial for organizations to respond to market opportunities swiftly and remain competitive. Software is also becoming increasingly more complex. Organizations seek to improve both the effectiveness and the efficiency of software engineering to enable further growth without increasing overhead and losing flexibility.

Continuous software engineering is a paradigm aiming to streamline software engineering by delivering software frequently and in small increments, and by doing so reaping different benefits from fast customer feedback, to continuous value delivery to said customers. Smaller increments are potentially easier to plan, develop, integrate, and verify. On the customers' side, more minor updates ought to create less disruption and are easier to adopt in contrast to big-bang software updates requiring downtime and catching up with the new features. 
Software vendors can collect more focused telemetry and customer feedback to steer further product development~\cite{humble2018accelerate}. More frequent and smaller software updates enable customers to provide more focused feedback empowering them to participate in steering the product more actively. In turn, this allows collaborative and experience-based business models~\cite{humble2018accelerate}. 

Continuous integration and delivery (CI/CD) is part of continuous software engineering. We consider how organizations can apply continuous principles throughout the whole software engineering process throughout inception, development, integration, verification, delivery, operation, use, collection of feedback, and planning the next software iteration steps~\cite{fitzgerald2017continuous}

However, few organizations have adopted continuous engineering beyond automating tests and other repetitive development tasks. To the best of our knowledge, an industrial-scale end-to-end pipeline from inception of a feature to collecting and analyzing customer feedback has yet to be demonstrated in peer-reviewed literature~\cite{shahin2017continuous}.

The potential benefits of continuous software engineering have gained much attention. Benefits like flexibility, efficiency, and improved time-to-market are appealing to most companies. However, the ability to retrofit an existing organization, what parameters determine the suitability, and what trade-offs are associated with adopting continuous software engineering remain largely unexplored.

This invited position paper discusses the potential, and highlights challenges hindering the widespread adoption of continuous software engineering. This is a position paper based on ongoing research with a dozen companies from multiple domains, featuring different market positions and customers, as part of the KKS research profile project Software Engineering Rethought (see \url{http://rethought.se}). 

Overall, we observe that end-to-end continuous software engineering, as per many recommendations, may not be applicable everywhere~\cite{humble2018accelerate,fitzgerald2017continuous}. However, organizations can adopt parts of the pipeline to streamline software engineering~\cite{klotins2021costsbenefits}. The challenge lies in the critical evaluation of the current situation and the goals of the company to maximize the gains from adopting continuous software engineering principles. 

This paper is structured as follows: Section 2 presents the potential of an end-to-end continuous software engineering. In section 3, we discuss challenges and the need for further research, before concluding our paper in section 4.


\section{Continuous Software Engineering in a Nutshell}

The idea of continuous software engineering originates from lean principles in manufacturing. One of the key principles in lean is to reduce waste and maximize customer value by implementing the flow. That is, linking all relevant production steps together and minimizing the lead time of each step. In software engineering, this principle can be implemented by delivering small increments of software~\cite{poppendieck2011principles}.

To enable the development and delivery of small software increments, software vendors need to implement a software delivery pipeline. The pipeline picks up the latest changes in the source code and automatically performs testing, integration, delivery, and other required steps to make the latest changes available to end-users. Once end-users start using the software, feedback and telemetry is relayed back to the software vendor for analysis and decision support to steer further product decisions.

In an idealized scenario, end-users gain access to the latest features and start generating feedback minutes after developers have finished the development~\cite{fitzgerald2017continuous,humble2018accelerate}.

\begin{figure}
\includegraphics[width=\textwidth]{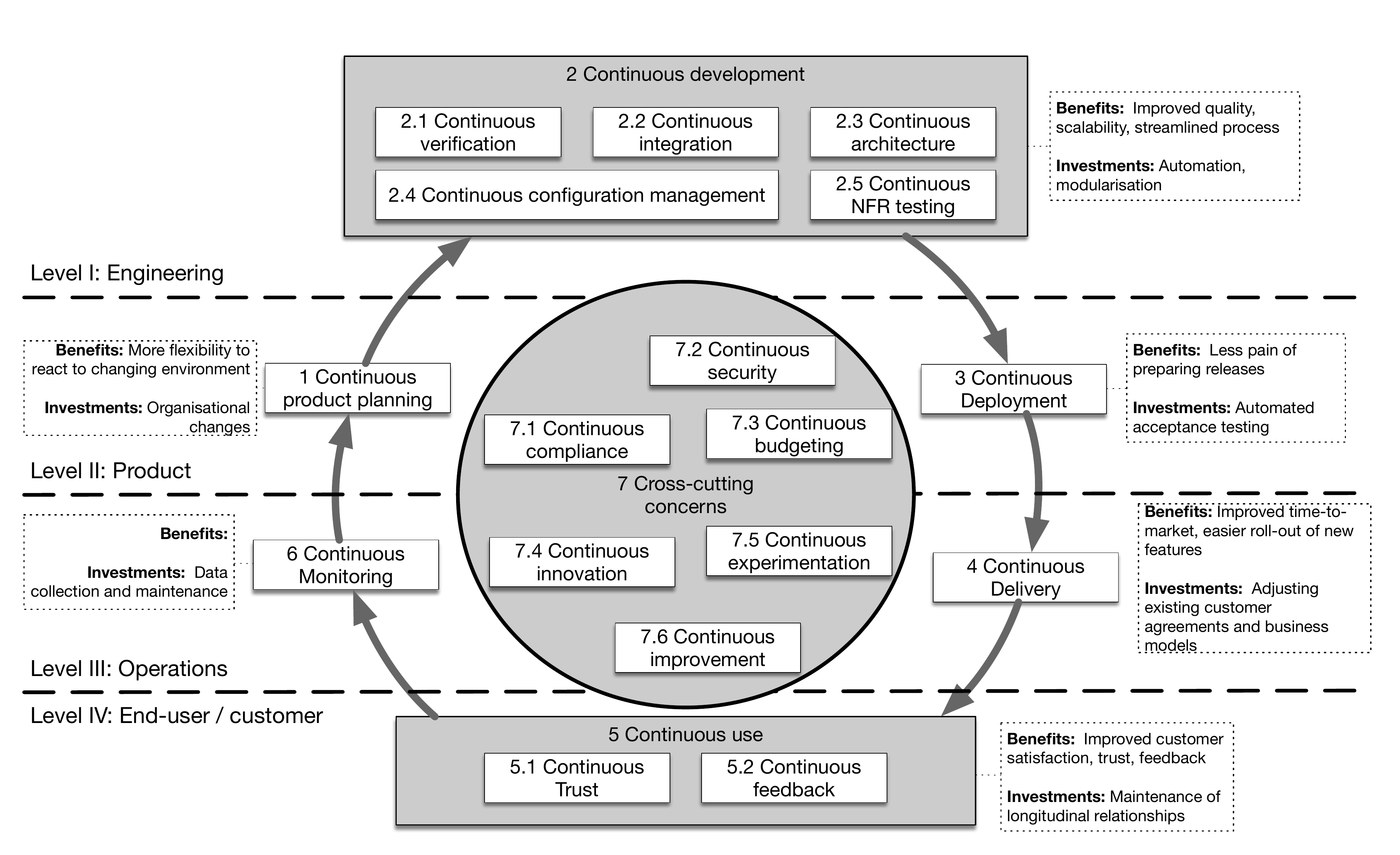}
\caption{An overview of state-of-the-art continuous software engineering pipeline} \label{fig:model}
\end{figure}

The state-of-the-art view on the end-to-end pipeline, along with the key steps, is shown in Fig,~\ref{fig:model}. The figure shows an idealized scenario that requires adaptations to fit any real-life scenario. 

In the figure, we show the key steps of the pipeline denoted with rectangular boxes. The arrows represent the flow of software and related artifacts through the pipeline. Dashed lines represent levels of stakeholders involved in product development. 

We differentiate between four levels of stakeholders, present how continuous software engineering affects each level, and present several potential  opportunities (\textbf{PT}) associated with adopting continuous engineering. The PT's are idealized positive outcomes that can be a result of realizing good continuous practices, used here to highlight potential benefits only for illustration purposes.  






\paragraph{Level I: Engineering organization} receives plans from the product planning organization and turns these plans into working software for deployment (Step 3). The engineering organization implements build, test, integration, and deployment automation; ensures the organization and software architecture supports incremental and parallel work. Note that the exact engineering activities are case-specific. In the figure, we illustrate the most common activities.

\textbf{PT 1:} Automation and parallel work on small increments are shown to improve efficiency quality, reduce stress and improve developer satisfaction. Independent, cross-functional teams taking responsibility for specific features and modular architectures allow scaling of development organization with minimal need for additional overhead.

\paragraph{Level II: Product organization} uses various inputs to devise plans for further product development (Step 2). These plans guide the engineering organization. Once the engineering is complete, the engineering organization returns working software (Step 3). 

\textbf{PT 2:} Working with lightweight plans and quick turnaround time allows product organization to rapidly adjust and explore new market opportunities. In dynamic markets, extensive market research and analysis could be counterproductive as the results are already outdated when they arrive. Instead, organizations could adopt a more experiment-driven approach and try out new ideas. It is possible if testing new ideas and recovering from unsuccessful experiments are reasonable effort and risk.

\paragraph{Level III: Operations} takes the working software from the product organization and makes it available for the end-users (Step 4). The working software generates feedback and telemetry, enabling monitoring of customer behavior (Step 6) and input for further planning (Step 1).

\textbf{PT 3:} Automating deliveries and frequently delivering to customers helps improve time-to-market and reduces release pains. Release pain is associated with developers required to put in extra effort to make sure the upcoming release is ready for customers. For large releases, it is associated with additional stress and workload. However, when releasing small incremental updates, the stress and workload associated with each release are minimized.

\textbf{PT 4:} Implementing telemetry and automated collection of feedback allows monitoring the software in use and making adjustments as needed. Quick turnaround time allows rapid release of patches and bug fixes, thus reducing the adverse effects of slipped defects. From the customers perspective, automated, seamless updates remove the need to update the software manually.

\paragraph{Level IV: End-user/customer organization} receives and uses the software (Step 5). Frequent upgrades and continuous access to new features enable service and experience-oriented business models. Such models encourage trust (Step 5.1), and the end-users are incentivised to actively participate in the product development by providing feedback (Step 5.2). 

\textbf{PT 5: } Both customers and the vendor can benefit from more closer experience and service based collaboration model. It facilitated trust and created incentives for customers not to switch to another vendor.

\paragraph{Cross-cutting concerns} transcends organizational levels. For instance, continuous improvement (Concern 7.6) attempts to measure and fine-tune the whole cycle continuously. Continuous experimentation allows a product organization to set up quick experiments to test market responses to new ideas (Concern 7.5).

\textbf{PT 6: } Continuous collection of data and frequent execution of the delivery process allows to systematically and continuously improve the process with every iteration.

\textbf{PT 7: } Transparency in planning and road maps of the development company can achieve a number of benefits. (i) Internal coordination in the development organization of what is in the pipe short- and medium term. (ii) Customers can be prepared of what is coming to prepare and plan for changes and benefits. (iii) Customers and the development organization can rather via transparency act as partners as interested customers can be active in the planning and release work.

In summary, adopting an end-to-end continuous software delivery pipeline as shown in Fig.~\ref{fig:model} allows the software vendor to increase internal efficiency and streamline value creation.

\section{Challenges and Future Needs}

Despite many potential advantages, the utilization of continuous software engineering remains scarce relative to plan-driven or sprint-based engineering models. Several of our partner companies have started adopting initiatives moving towards continuous software engineering. However, the adoption is far from straightforward and hindered by many challenges. In this section, we present challenges that, in our view and experience, should be solved to support an industry-wide adoption of continuous engineering principles.

\paragraph{Challenge 0: Should you do it?} The tendency to go for one-size-fits-all solutions and the power of ``the next thing everyone is doing'' can cause as much damage as benefit. A key is to start with an analysis of both the capacity and goals of the development organization, as well as the capacity and needs of the customer's organizations. 

The development organization needs to specify goals and to break down these goals to a level where success (or failure) to achieve these goals can be "measured" by the organization, continuously as changes are realized. This is paramount to a risk analysis, and also a cost benefit analysis anchored in reality and objective observations. A significant bonus is that this activity also enables creating a common understanding of the potential, implications and direction of the changes from items impacting individual teams, to entire departments and the whole organization, as well as how they are dependent and tie together. In this, it is important to separate between the treatment (change/tool/action/new way of working) and the goals and attainment of the goals to enable an evidence-based change continuously.

At the same time, knowing your customer and the impact of continuous on them and the relationship with them is critical. What is the benefit for the customer? Is the domain and agreements between the customer and the development company suitable for a continuous environment? One specific example could be if the customer wants continuous changes in the product, and is there a telemetrics/data feedback system in place to report on new features and changes enabling feedback? In some domains and for some development organizations a continuous model is non-controversial, but in some other domains most of the benefits attained via try-and-learn-improve are impossible to achieve. For example, an organization and customer operating in (even partly) safety-critical products, or where down-time of products incur substantial consequences, the benefit/risk/cost calculation is significantly different. 

Significant research needs to be conducted, and usable and useful models need to be developed, to establishing cost-effective ways to continuously evaluate and course correct the continuous changes needed to achieve a continuous product development environment; maybe inspired by a hierarchically connected Goal-Question-Metric model~\cite{khurum2013software} - back-filled by metrics and data collected via the continuous feedback cycle. As such, a goal and measurement program is created (and scaled to be usable and useful for the organization in question) and it is important to realize that it might lead to the conclusion that continuous engineering in its entirety and complete idealized form might not fit your organization. However, there are probably many parts of continuous engineering that are beneficial for most organizations. Start there.

\paragraph{Challenge 1: Determining adoption goals and constraints.} We observe that organizations often put forward aims like improving speed and efficiency to drive the adoption of continuous engineering. However, such aims are too vague to be measured and drive systematic improvements. When interpreted by different parts of the organization, vague goals may come at odds, or even worse result in different interpretations of said goals resulting in different direction of work and sub-optimizations. 

For example, an R\&D unit could interpret the efficiency as maximizing the delivery of new experimental features. For operations, efficiency could mean minimizing resources to ensure services availability. Without a joint view of what efficiency means and how it is measured in the given organizational context, internal deadlocks may arise, hindering the company's adoption effort and optimal operation.

Attaining specific goals often imply trade-offs. As in the earlier example, it could be challenging to launch many innovative features and reach high stability of services simultaneously. Such trade-offs need to be identified and analyzed to understand the associated constraints and degrees of freedom. 

The analysis of organizational goals, constraints, and trade-offs should drive the organizational change towards continuous software engineering. One powerful tool in coordinating an organization, and making goals clear is to break-down goals into measurable effects or metrics. This requires a number of steps. The terms  "value", "efficiency", or "effectiveness" need to be defined for each part of the organization~\cite{khurum2013software}. These definitions need to be coordinated and streamlined - in essence shared. Then, metrics on how to ascertain level of success have to be detailed very early. How to measure if you are improving towards a goal is a prerequisite before adopting any treatment or change. For example, if you say a specific practice or set of practices should be realized to improve customer value, what type of customer value, and how do you measure it?  This analysis would also pinpoint bottlenecks, inefficiencies, and areas of improvement, in addition to acting as a coordinating force as an organization realizes changes. You need to be able to measure benefits of a change as well as you measure the cost of said change. 

\paragraph{Challenge 2: Considering the return-of-investment perspective} Retrofitting an organization with a new continuous engineering pipeline and new ways of working is a substantial investment. The investments should be justified with potential benefits and be aligned with organizational objectives (Challenge 1). Notably, the organization must be prepared to realize the potential and materialize the benefits. 

For example, an organization may invest in data collection (PT 5) and gain the potential of improved data-driven decisions (PT 2). However, if the rest of the organization is not ready to use the data in decision support, the potential is not realized, and the investment is wasted. Furthermore, customers may be slow in adopting new features, thus delaying the feedback and nullifying its value to the organization.

The organization should perform a cost-benefit analysis to gauge the viability of any goals, new practices, and working methods. Adoption champions should do such analysis in parallel with determining adoption goals, trade-offs, and constraints, see Challenge 1.

To address Challenges 1-2, we propose a model supporting inventory of goals, trade-offs, and return-of-investment calculation to support the systematic adoption of continuous practices. As a part of this analysis, the break-down of goals into defining the terms used to establish a common vocabulary, then establishing how to "measure" goal attainment by breaking down goals into measurable items. This also allows for course corrections during the changes associated with retrofitting.

\paragraph{Challenge 3: Focus on cost savings instead of value and potential creation.} Our industry partners often mention the need to reduce cost and improve efficiency (PT 1) as reasons for considering continuous software engineering. Potential benefits like new value streams, improved time-to-market, and new business models are rarely, if at all, mentioned.

The question if adopting continuous practices would provide some cost savings is flawed. We compare such a question to asking whether eating healthier will be cheaper.Both eating healthier and adopting continuous engineering will likely cost more. Developing tests, maintaining test suites and automation infrastructure, collecting and maintaining test data, refactoring software architectures, and driving organizational adjustments will push the overall cost of software engineering upwards. At the same time, speed and flexibility will open up opportunities for new offerings enabling the organization to become faster at responding to market opportunities, among other benefits. Eating healthier may cost more, but may enable you a better and longer life.

We propose emphasizing the business value arising from streamlining software delivery over the exclusive focus on potential cost savings. More cross-disciplinary studies are needed to explore the business value perspective. We observe that continuous shares many characteristics with so called "digital transformation". Considering how continuous engineering fits into the broader organizational transformation  could help to shift the focus from cost savings to unlocking the organizational potential~\cite{klotins2022briding}. 

\paragraph{Challenge 4: Remembering Conway's law.} In 1967, M. Conway formulated an adage that organizations develop systems that mimic their communication structure. In software engineering terms, software architecture should follow organizational structures. To change the prior, one needs to change the latter first. 

Best practices of continuous software engineering dictate that development work should be done by real cross-functional teams that own the development of whole features. A team should have full responsibility from generating improvement ideas to development, testing, delivery, and telemetry analysis. This practice allows maintaining a modular architecture and to minimize the gap between organizational structures and software architectures.

From discussions with our partner companies, we learned that organizations often find their software architectures monolithic and poorly suited for parallel development, automation, and modular deployments. The software travels through various organizational silos, each performing a specific function without the complete picture. Handover from one silo to another creates friction and bottlenecks.

Attempts to break the software monolith often lead to failure as the surrounding organization remains the same. Changing the organizational structures is often extremely challenging due to internal inertia and resistance to changes.

We propose to explore the adoption of continuous engineering practices from the organizational view first. That is, analyze what organization silos and bottlenecks currently limit the continuous flow of software. Once organizational inefficiencies are addressed, the software delivery process can be further improved with tools such as automation. In addition, the detailed goals and following follow-up (metrics) should be owned by the parts of the organization (the teams) that own the module or part of the architecture. This way you can measure the fit of the organization to the goals of development. For example, a change in organizational structure might result in (metric) less waiting times between teams.

\paragraph{Challenge 5: Ever-increasing complexity}

Contemporary market-driven software engineering exists in a dynamic environment. It faces mercurial market influences, changing organizational goals, technologies, and the growing size and complexity of software and the surrounding organization.

A single person, or a small group of people, can no longer grasp the complexity and make optimal and timely decisions using their expertise alone. This increase in complexity has far-reaching implications for how organizations make decisions. 

One viable way forward could be to consider data collection and analysis as an integral part of the product and the engineering process. The organization can use data to support decisions both on what features to develop (\textit{What to build?}) and how to improve the engineering process (\textit{How to build?}). 

Furthermore, retrospective analysis, that is, analyzing past events, has limited use in an increasingly changing environment. There is a potential to explore the applicability of inferential statistical methods, simulations, and machine learning techniques to make relevant predictions. This is however subject to significant research initiative to develop transparent tools and methods that can be trusted by practitioners to predict and simulate developments, at least in the short-term.

\section {Conclusions}
This invited position paper analyzes continuous software engineering and pinpoints several challenges of adopting ongoing software engineering from our collaboration with industrial partners. The challenges emphasize the need for systematic methods to analyze organizational goals, context, structures, constraints, among other contextual factors, to remove inefficiencies and realize the full potential of software-intensive products and services. 

We wish to highlight that most inefficiencies and obstacles to streamlined value delivery can be traced to the lack of organizational alignment and coordination. The spirit of continuous engineering is to identify and remove such hindrances systematically. In the end, continuous software engineering is more about being pragmatic and disciplined in engineering than it is about automation.

\bibliographystyle{splncs04}
\bibliography{bibliography.bib} 




\end{document}